\documentclass[aps,pra,twocolumn,groupedaddress,showpacs,preprintnumbers,
amsmath,amssymb]{revtex4}
\usepackage{graphicx}
\usepackage{dcolumn}
\usepackage{bm}
\usepackage{amsmath}

\newcommand{\eb}{\bar{E}}
\newcommand{\et}{\widetilde{E}} 
\newcommand{\ob}{\bar{\Omega}}
\newcommand{\ot}{\widetilde{\Omega}} 
\newcommand{\rhob}{\bar{g}}
\newcommand{\rhot}{\widetilde{g}} 
\newcommand{\nc}{{\cal N}}
\newcommand{\ncb}{\bar{\cal N}} 
\newcommand{\nct}{\widetilde{\cal N}}
\newcommand{\mub}{\bar{\mu}} 
\newcommand{\mut}{\widetilde{\mu}}

\begin{document}

\title{Average ground-state energy of finite Fermi systems}

\author{M. Centelles$^1$, P. Leboeuf$^2$, A. G. Monastra$^3$, J.
Roccia$^2$, P. Schuck$^4$, and X. Vi\~nas$^1$}

\affiliation{
$^1${\it Departament d'Estructura i Constituents de la Mat\`eria,
Facultat de F\'{\i}sica,
Universitat de Barcelona,
Diagonal {\sl 647}, {\sl 08028} Barcelona, Spain}
\\
$^2${\it Laboratoire de Physique Th\'eorique et Mod\`eles Statistiques,
CNRS, Universit\'e de Paris--Sud, UMR {\sl 8626}, {\sl 91405} Orsay
cedex, France} \\
$^3${\it TU Dresden Institut f\"ur Theoretische Physik, {\sl 01062}
Dresden, Germany} 
\\
$^4${\it Institut de Physique Nucl\'eaire, IN{\sl 2}P{\sl 3}--CNRS,
Universit\'e de Paris--Sud, {\sl 91406} Orsay cedex, France}
}


\begin{abstract}
Semiclassical theories like the Thomas-Fermi and Wigner-Kirkwood methods give
a good description of the smooth average part of the total energy of a Fermi
gas in some external potential when the chemical potential is varied. However,
in systems with a fixed number of particles $N$, these methods overbind the
actual average of the quantum energy as $N$ is varied. We describe a theory
that accounts for this effect. Numerical illustrations are discussed for
fermions trapped in a harmonic oscillator potential and in a hard wall cavity,
and for self-consistent calculations of atomic nuclei. In the latter case, the
influence of deformations on the average behavior of the energy is also
considered.
\end{abstract}


\maketitle

\section{Introduction}

A basic problem in the physics of finite fermion systems such as, e.g., atoms,
nuclei, helium clusters, metal clusters, or semiconductor quantum dots, is the
determination of the ground-state energy $E$. A standard decomposition, deeply
rooted in the connection of classical and quantum physics, is to write $E$ as
the sum of an average energy $\eb$ and a fluctuating part $\et$ 
\cite{BB,RS,Strut}:
\begin{equation}
E (N) = \eb (N) + \et (N) \ .
\label{eq1} \end{equation}
The largest contribution, $\eb$, is a smooth function of the number $N$ of
fermions. The shell correction $\et$ has a pure quantal origin and displays,
instead, an oscillatory behavior as a function of $N$.

Equation (\ref{eq1}) underlies the usefulness of the so-called mass formulae,
like the liquid drop model for nuclei or for metal clusters, of which the
oldest example is the well-known Bethe-Von Weizs\"acker mass formula for the
binding energy of nuclei. The decomposition (\ref{eq1}) is also at the basis
of semiclassical and statistical techniques that are used to investigate how
the properties of global character of fermion systems vary with the particle
number $N$. Such is the case for instance of the celebrated Thomas-Fermi and
Wigner-Kirkwood theories \cite{BB,RS}. These methods often provide deep
physical insights which may be otherwise obscured behind a full quantum
calculation.

It is recognized, however, that the semiclassical calculations of $\eb (N)$
for fermion systems in either external potentials or self-consistent mean
fields show systematic deviations with respect to the actual average of the
exact quantum results \cite{BB,RS,March,BCKT,PS,CVBS,cent06}. For example, in
spherically symmetric calculations one finds that, as a function of the number
$N$ of particles, the difference $E (N) - \eb (N)$ between the (fluctuating)
exact value $E (N)$ and the (smooth) semiclassical average $\eb (N)$ {\em does
not\/} oscillate around zero. In general, for fermions in a fixed external
potential, semiclassical calculations overbind the true average of the quantum
energy. One of our purposes in the present work is to explain the origin of
this effect, and to derive an explicit formula that allows to compute the
correct average behavior of $E (N)$ in fermion systems. Related studies are
the works of Refs.\ \cite{ivanyuk,pomorski} where a particle number conserving
shell correction method has been pursued.

Additional contributions to the average part of the ground-state energy come
in fact from a careful analysis of the oscillatory term $\et (N)$. Because
this fluctuating function is evaluated at discrete values of the chemical
potential (that correspond to integer values of the particle number), its
average value is generically non-zero and therefore contributes to the average
part of $E(N)$. This phenomenon is related to the different physical
descriptions of quantum mechanical systems obtained from different
thermodynamic ensembles, the grand canonical and the canonical in the present
context. This subtle topic has played, in recent years, a crucial role in
understanding the physics of, e.g., persistent currents in mesoscopic metallic
rings \cite{agi}, or in trapped Bose-Einstein condensates \cite{gh}.

Our results are illustrated with two schematic models. Namely, we study the
average of $E(N)$ for fermions in a harmonic oscillator (HO) potential, via
the semiclassical Wigner-Kirkwood (WK) theory \cite{WK}, and for fermions in a
spherical cavity with sharp boundaries, via the Weyl expansion \cite{bh}. In
the former case, analytical expressions are available. Finally, and in
contrast to the previous examples where the confining potential is fixed, we
consider the influence of deformations and self-consistency on the average
behavior of $E(N)$, as well as other related topics. We find that for
self-consistent potentials with deformation degrees of freedom the behavior of
the average energy is qualitatively different.

\newpage

\section{Smooth behavior: grand canonical versus canonical ensembles}

The usual computation of the different terms in Eq.\ (\ref{eq1}) is as follows.
The single-particle level density $g (\varepsilon) = {\rm
Tr}\/ [\delta (\varepsilon - \hat{H}) ]$ of a quantum fermion system can be
expressed as \cite{WK,BB,RS}
\begin{equation}
g(\varepsilon) = \frac{2}{(2 \pi \hbar)^3} \int \!\! \int \frac{\partial
 f_{\varepsilon}(\vec{r},\vec{p})} {\partial \varepsilon} 
 \, d\vec{p} \, d\vec{r}
\label{eq14}
\end{equation}
in terms of the phase space Wigner function $f_{\varepsilon}
(\vec{r},\vec{p})$. We have included a factor 2 to account for spin
degeneracy. Then, for a set of fermions in a potential well filled up to an
energy $\mu$, the number of states (accumulated level density) and the
ground-state energy are obtained from $g(\varepsilon)$ through
\begin{equation}
\nc (\mu) = \int^{\mu}_0 g(\varepsilon) \, d\varepsilon \ , \qquad E(\mu) =
\int^{\mu}_0 \varepsilon \, g(\varepsilon) \, d\varepsilon \ .
\label{eq13} \end{equation}
Inserting the Wigner-Kirkwood expansion of the Wigner function
$f_{\varepsilon}(\vec{r},\vec{p})$ in powers of $\hbar$ in Eq.\ (\ref{eq14})
produces a smooth function $\rhob (\varepsilon)$, where the leading order
gives rise to the Thomas-Fermi term. This procedure is well documented in the
literature \cite{BB,RS,BP,BP2,BP3,BP4}. Inserting the latter series for
$g(\varepsilon) \approx \rhob (\varepsilon)$ into Eq.~(\ref{eq13}) yields the
semiclassical $\hbar$ expansions for $\ncb (\mu)$ and $\eb (\mu)$.
Alternatively, for a Fermi gas contained in a hard wall cavity, one inserts in
Eqs.\ (\ref{eq13}) the corresponding Weyl expansion \cite{bh} of the average
single-particle level density $\rhob (\varepsilon)$. In both cases, Eqs.\
(\ref{eq13}) produce a series in decreasing powers of $\mu$ whose coefficients
depend on the shape of the potential.

These expressions provide in general accurate descriptions of the average
behavior of $g(\varepsilon)$, $\nc (\mu)$, and $E (\mu)$. For instance, for an
isotropic three dimensional HO potential of frequency $\omega$ one obtains the
well-known WK expressions \cite{BB,RS,BP,BP2,BP3,BP4}
\begin{eqnarray}
\rhob(\varepsilon) & = &
\bigg[ \bigg(\frac{\varepsilon}{\hbar \omega}\bigg)^2 
       - \frac{1}{4} \bigg] \frac{1}{\hbar \omega}
+ \frac{17}{960} \, \hbar \omega \, \delta^{\,\prime}(\varepsilon) \,,
 \label{eq19}
\\
\ncb (\mu) & = & \frac{1}{3} \bigg(\frac{\mu}{\hbar \omega}\bigg)^3 -
\frac{1}{4} \frac{\mu}{\hbar \omega} \, ,
\label{eq20} 
\\ \eb (\mu) & = & \bigg[ \frac{1}{4} \bigg(\frac{\mu}{\hbar \omega}\bigg)^4 -
\frac{1}{8} \bigg(\frac{\mu}{\hbar \omega}\bigg)^2 - \frac{17}{960} \bigg]
\hbar \omega \, .
\label{eq21} 
\end{eqnarray}
The last term in Eq.\ (\ref{eq19}) contains the derivative of the delta
function $\delta(\varepsilon)$. This term and the last term in Eq.\
(\ref{eq21}) stem from the corrections of order $\hbar^4$ to $\rhob$ and
$\eb$, respectively. In the HO potential the $\hbar^4$ contribution to $\ncb$
vanishes.

\begin{figure}
\includegraphics[width=0.90\columnwidth,clip=true]{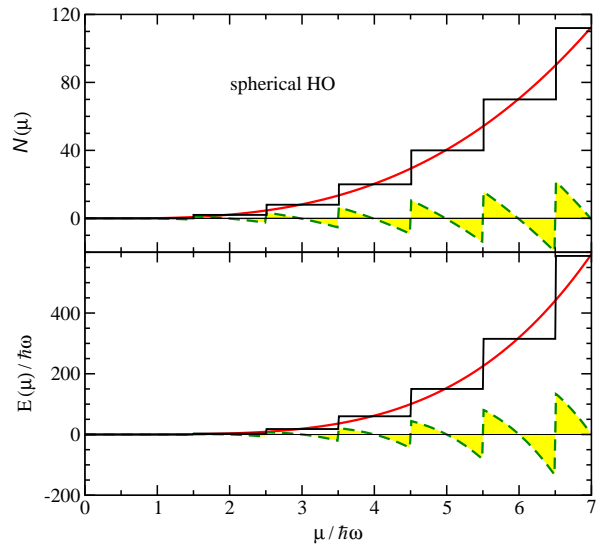}
\caption{\label{figure1} Accumulated level density (upper panel) and total
energy (lower panel) with spin-degeneracy 2 for a spherical HO potential as a
function of chemical potential $\mu$. Staircase, solid, and dashed lines
correspond to the quantum, semiclassical WK, and shell correction (quantum
minus semiclassical) values, respectively.}
\end{figure}

Figure \ref{figure1} displays the comparison between the exact quantum
mechanical quantities and the smooth approximations
(\ref{eq20})--(\ref{eq21}). The upper panel shows the accumulated level
density $\nc (\mu)$ for a set of fermions in a spherical HO potential, as a
function of $\mu/ \hbar\omega$. The quantum result exhibits discontinuities at
each major shell ($N= 2$, 8, 20, 40, 70, 112 in the present case) and is
represented by a staircase function which fluctuates around the smooth WK
curve provided by Eq.\ (\ref{eq20}). The oscillatory part of $\nc (\mu)$
(dashed curve) contains the fluctuations due to shell effects. They are seen
to oscillate around zero, with a vanishing net average, as $\mu$ is varied.
The lower panel of Fig.~\ref{figure1} displays the ground state energy
$E(\mu)/\hbar \omega$ for the same potential \cite{footnote}. Again, the
smooth WK curve excellently averages the quantum result and the shell energy
fluctuates around zero.

The fact that the average behavior of the remaining shell corrections is zero
for $E (\mu)$ and $\nc (\mu)$ can be explicitly checked. The general
semiclassical theory expresses the fluctuating parts $\et (\mu) = E(\mu)-\eb
(\mu)$ and $\nct (\mu) = \nc (\mu) - \ncb (\mu)$ as sums over the classical
periodic orbits of the system at energy $\mu$ \cite{bab,gutz,BB,seville}. Each
term in $\et (\mu)$ and $\nct (\mu)$ is an oscillatory function of the
chemical potential (through the action of the corresponding orbit), whose
average over a chemical potential window is zero. In the particular case of
the HO potential the semiclassical approximation turns out to be exact (see
e.g.\ \cite{BB,BJ}), and takes the form
\begin{eqnarray}
\nc (\mu) \!\! &=& \!\! \ncb (\mu) + 2 \sum_{M=1}^{\infty}
 \frac{(-1)^M}{M_r^3} \big[ \mu_r \cos \mu_r \nonumber \\ & & \mbox{} + \big(
 \mu_r^2 - 2 - \textstyle{\frac{1}{4}} M_r^2 \big) \sin \mu_r \big] ,
\label{eq23}
\\ E(\mu) \!\! &=& \!\! \eb (\mu) + 2 \sum_{M=1}^{\infty} \frac{(-1)^M}{M_r^4}
\big[ \big( 3 \mu_r^2 - 6 - \textstyle{\frac{1}{4}} M_r^2 \big) \cos \mu_r
\nonumber \\ & & \mbox{} + \big( \mu_r^2 - 6 - \textstyle{\frac{1}{4}} M_r^2
\big) \mu_r \sin \mu_r \big] \hbar \omega ,
\label{eq24} \end{eqnarray}
where $M_r \equiv 2 \pi M$, $\mu_r \equiv M_r \, \mu/ \hbar \omega$, and $\ncb
(\mu)$ and $\eb (\mu)$ are given by Eqs.\ (\ref{eq20}) and (\ref{eq21}),
respectively. These expressions illustrate explicitly that the average of the
fluctuating parts $\et (\mu)$ and $\nct (\mu)$ over a chemical potential
window is zero, $\langle \et (\mu) \rangle_\mu = \langle \nct (\mu)
\rangle_\mu = 0$.

In real physical Fermi gases with a well defined number of particles the
various quantities, like masses or many-body level densities, are not studied
as a function of the chemical potential $\mu$ but rather as a function of the
particle number $N$. For instance, the ground-state energy $E(N)$ of the
system consists in the sum of the single-particle energies of the $N$ lowest
single-particle states (taking into account spin-degeneracy). Thus Eqs.\
(\ref{eq23}) and (\ref{eq24}) are related to the grand canonical ensemble. The
qualitative behavior of the function $E (N)$ as a function of $N$ is in
general quite different from the behavior of $E (\mu)$.

Based on e.g.\ the Wigner-Kirkwood method, the usual way to calculate the
function $E (N)$ is as follows. Having determined, in that approximation, the
energy $\eb (\mu)$ and the accumulated level density $\ncb (\mu)$, one first
fixes the chemical potential (or Fermi energy) in terms of the particle-number
$N$ by inverting the function
\begin{equation} \label{106}
\ncb (\mub) = N \Longrightarrow \mub_{\scriptscriptstyle N} = \mub (N) \ .
\end{equation}
To be consistent in the notation, we use $\mub_{\scriptscriptstyle N}$ to
denote the value of the chemical potential for a given $N$ determined from the
WK (or Weyl) approximation, to stress that, in this approximation, $\mu$ is
computed by inverting the smooth part $\ncb$ and not the exact counting
function $\nc$. Finally, one obtains the smooth Wigner-Kirkwood or Weyl term
$\eb (N)$ replacing $\mu$ by $\mub_{\scriptscriptstyle N}$ in $\eb (\mu)$,
\begin{equation} \label{108}
\eb (N) = \eb (\mub_{\scriptscriptstyle N}) = \int^{\mub_{\scriptscriptstyle
N}}_0 \varepsilon \, \rhob (\varepsilon) \, d\varepsilon \ .
\end{equation}
For example, applying this procedure to the isotropic HO potential,
from the leading terms of Eqs.\ (\ref{eq20}) and (\ref{eq21}) one
straightforwardly obtains
\begin{equation} \label{eq25}
\eb (N) = \frac{1}{4} (3 N)^{4/3} \hbar\omega \ ,
\end{equation}
which is the leading-order Thomas-Fermi result. It shows that in a HO the
leading dependence of the average energy per particle, in units of
$\hbar\omega$, is $\propto N^{1/3}$.

\begin{figure}
\includegraphics[width=0.90\columnwidth,clip=true]{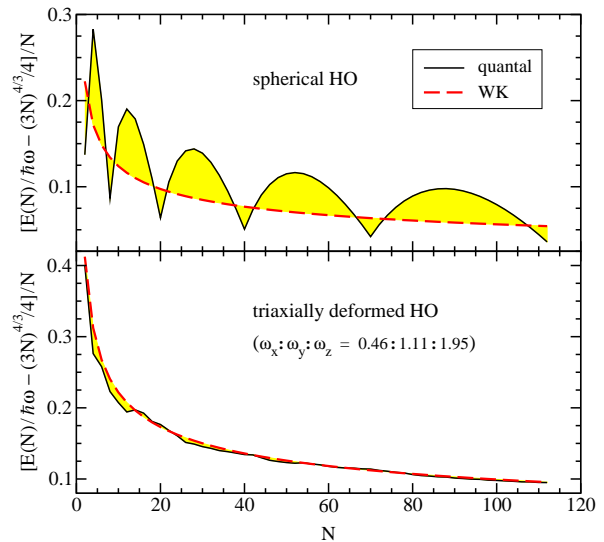}
\caption{\label{figure2} Upper panel: quantum and WK values of the
energy per particle for a spherical HO potential as a function of the
number of particles, in units of $\hbar \omega$. The leading
$N$-dependence given by Eq.\ (\ref{eq25}) is subtracted from {\em both}
curves. Lower panel: the same as in the upper panel but
for a strongly triaxially deformed HO potential. Notice that the WK
curves are different in the spherical and deformed cases.}
\end{figure}

The full Wigner-Kirkwood function $\eb (N)/N$ computed for the HO potential
including up to the fourth-order contributions in $\hbar$ is plotted in
Fig.~\ref{figure2} (dashed line) as a function of the particle number $N$, in
units of $\hbar\omega$. It is compared to the exact quantum result (solid
line). To better visualize the quantum oscillations with changing $N$, we have
subtracted the dominant $N^{1/3}$ dependence [recall Eq.\ (\ref{eq25})] from
{\em both} the quantum and the WK curves. The upper panel of
Fig.~\ref{figure2} displays the results for the isotropic HO potential. The
lower panel is for a strongly deformed potential and it will be discussed
later on. Focusing on the isotropic HO, one sees that, as expected, the
general trend of the smooth WK result turns out to be quite correct in
comparison with the global particle number dependence of the quantum energies.
There is, however, a systematic deviation in the sense that the WK curve does
not pass as a function of $N$ through the average of the quantum values. This
is clearly seen from the large asymmetry of the shaded regions above and below
the WK curve in the upper panel of Fig.~\ref{figure2}. One notices that the WK
result overbinds with respect to the true average of the quantum values when
$N$ is varied in the spherical symmetry. The same situation prevails in other
problems of atomic and nuclear physics as well as in self-consistent mean
field calculations \cite{BB,RS,March,BCKT,PS,CVBS,cent06}.

\section{Contribution of the oscillatory corrections}

The previous results show that the function $\eb (N)$ does not describe
appropriately the average behavior of $E (N)$. We now discuss the origin of
the discrepancy, and the way to correct it.

In systems with a well defined number of particles the chemical potential
$\mu$ takes discrete values. These values do not occur at random. For
instance, for an even number of particles and non-degenerate single-particle
states, a standard rule is to locate the chemical potential half-way between
the last occupied and the first unoccupied single-particle states. Fixing a
particular rule to determine the chemical potential at a given number of
particles introduces a bias in the sampling of the values of $\mu$ (with
respect to a uniform, random distribution of $\mu$). Due to this bias, when
the oscillatory part of the energy $\et (\mu)$ is evaluated over the set of
discrete points it produces, generically, a function whose average is
different from zero. To compute that average we proceed as follows.

The decomposition of the single-particle level density into a smooth part and
a fluctuating part,
$$
g (\varepsilon) = \rhob (\varepsilon) + \rhot (\varepsilon) \ ,
$$ 
where $\rhob (\varepsilon)$ is the WK (or Weyl) smooth part and $\rhot
(\varepsilon)$ is given by the sum over periodic orbits mentioned above
\cite{BB}, induces a corresponding decomposition of the integrated density 
[cf Eq.\ (\ref{eq13})]:
\begin{equation} \label{107}
\nc (\mu) = \ncb (\mu) + \nct (\mu) \ .
\end{equation}

For a given number of particles $N$, the chemical potential is defined by
inversion of the exact accumulated level density
\begin{equation} \label{110}
\nc (\mu) = N \Longrightarrow \mu_{\scriptscriptstyle N} = \mu (N) \ .
\end{equation}
As the particle number $N$ increases, it is natural to decompose the chemical
potential into smooth and fluctuating parts:
$$
\mu_{\scriptscriptstyle N} =
\mub_{\scriptscriptstyle N} + \mut_{\scriptscriptstyle N} \ .
$$ 
The average part $\mub_{\scriptscriptstyle N}$ satisfies Eq.\ (\ref{106}).
Assuming that $\mut_{\scriptscriptstyle N} \ll \mub_{\scriptscriptstyle N}$, a
Taylor expansion of the smooth part in powers of $\mut_{\scriptscriptstyle N}$
around $\mu = \mub_{\scriptscriptstyle N}$ in Eq.\ (\ref{107}) (no Taylor
expansion is allowed for the fluctuating term $\nct$, because it is not a
regular function) yields, to lowest order,
\begin{equation} \label{101}
\mut_{\scriptscriptstyle N} = - \frac{1}{\rhob} \ \nct
(\mu_{\scriptscriptstyle N}) \ ,
\end{equation}
where we denote 
\begin{equation} \label{eqg}
\rhob = \rhob (\mub_{\scriptscriptstyle N}) \ .
\end{equation}
Similarly, the energy may be decomposed as
\begin{equation} \label{109}
E(\mu) = \eb (\mu) + \et (\mu) \ .
\end{equation}
In a system with a well-defined number of particles, the smooth part $\eb (N)$
of the exact function $E (N) = E (\mu_{\scriptscriptstyle N})$ was defined in
(\ref{108}), $\eb (N) = \eb (\mub_{\scriptscriptstyle N}) =
\int^{\mub_{\scriptscriptstyle N}}_0 \varepsilon \, \rhob (\varepsilon) \,
d\varepsilon$. The fluctuating part is thus defined as
\begin{equation} \label{111}
\et (N) = E (N) - \eb (N) \ .
\end{equation}

In order to compute $\et (N)$, and in particular its average over some
particle-number window $\Delta N$ around $N$, it is convenient to express the
energy in terms of the grand potential $\Omega = - \int^{\mu} \nc
(\varepsilon) d \varepsilon$ using the thermodynamic relation
$$
 E (\mu) = \Omega (\mu) + \mu \, \nc (\mu) \ .
$$
Recalling the definition of $\mu_{\scriptscriptstyle N}$ and
$\mub_{\scriptscriptstyle N}$ [Eqs.\ (\ref{106}) and (\ref{110})], Eq.\
(\ref{111}) may be written
$$ \et (N) = \Omega (\mu_{\scriptscriptstyle N}) - \ob
(\mub_{\scriptscriptstyle N}) + \mut_{\scriptscriptstyle N} N \ ,
$$
where $\ob (\mub_{\scriptscriptstyle N}) = - \int^{\mub_{\scriptscriptstyle
N}} \ncb (\varepsilon) d \varepsilon$. Decomposing $\Omega
(\mu_{\scriptscriptstyle N})$ into its average and fluctuating parts,
expanding $\ob (\mu_{\scriptscriptstyle N})$ around $\mub_{\scriptscriptstyle
N}$ to second order in $\mut_{\scriptscriptstyle N}$, and using the
thermodynamic relations
$$
\frac{\partial \ob
(\mub_{\scriptscriptstyle N})}{\partial \mub_{\scriptscriptstyle N}} =
- \ncb  (\mub_{\scriptscriptstyle N}) 
\quad \mbox{and} \quad
\frac{\partial^2 \ob (\mub_{\scriptscriptstyle
 N})}{\partial \mub^2_{\scriptscriptstyle N}} = - \rhob \ ,
$$
we get
$$ \et (N) = \ot (\mu_{\scriptscriptstyle N}) - \rhob \
\mut_{\scriptscriptstyle N}^{2} /2 \ .
$$
Using Eq.\ (\ref{101}), this takes the form
\begin{equation} \label{102}
\et (N) = \ot (\mu_{\scriptscriptstyle N}) - \frac{1}{2 \ \rhob} \ \nct^2
(\mu_{\scriptscriptstyle N}) + {\cal O} (\mut_{\scriptscriptstyle N}^3) \ .
\end{equation}
Equation (\ref{102}) connects the fluctuations of the grand potential
(grand-canonical ensemble) to those of the energy at a fixed number of
particles (canonical ensemble). This connection, to lowest order, has been
exploited in recent years to analyze nuclear-mass fluctuations \cite{bl,oabl}.

One may be tempted to think that the average of $\et (N)$ over some
particle-number window $\Delta N$ around $N$, denoted $\langle \et (N)
\rangle_{\scriptscriptstyle N}$, is proportional, from Eq.\ (\ref{102}), to the
variance $\langle \nct^2 (\mu_{\scriptscriptstyle N})
\rangle_{\scriptscriptstyle N}$ of $\nct (\mu_{\scriptscriptstyle N})$, and
that the average of $E (N)$,
\begin{equation} \label{eav}
\langle E (N) \rangle_{\scriptscriptstyle N} = \eb (N) + \langle \et (N)
\rangle_{\scriptscriptstyle N} \ ,
\end{equation}
is thus lowered with respect to $\eb (N)$ (due to the minus sign in front of
$\nct^2 (\mu_{\scriptscriptstyle N})$ in Eq.(\ref{102})). However this is
wrong because, for the same reasons as for $\et (N)$, the average $\langle \ot
(\mu_{\scriptscriptstyle N}) \rangle_{\scriptscriptstyle N} \neq 0$. A
detailed calculation (cf the Appendix) shows that
\begin{equation} \label{103}
\langle \ot (\mu_{\scriptscriptstyle N}) \rangle_{\scriptscriptstyle N} =
\frac{1}{\rhob} \ \langle \nct^2 (\mu_{\scriptscriptstyle N})
\rangle_{\scriptscriptstyle N} +\frac{1}{8} \ \rhob \ \langle
s_{\scriptscriptstyle N}^2 \rangle_{\scriptscriptstyle N} - \frac{1}{6 \ \rhob} \ ,
\end{equation}
where $\langle s_{\scriptscriptstyle N}^2 \rangle_{\scriptscriptstyle N}$ is
the variance of the spacing $s_{\scriptscriptstyle N} =
\varepsilon_{\scriptscriptstyle N+1} - \varepsilon_{\scriptscriptstyle N}$
between two consecutive single-particle levels around $\mu_{\scriptscriptstyle
N}$. Taking the average with respect to the discrete points
$\mu_{\scriptscriptstyle N}$ in Eq.(\ref{102}), using Eq.\ (\ref{103}) for the
average of $\ot (\mu_{\scriptscriptstyle N})$ and expressing, for convenience,
the average $\langle \nct^2 (\mu_{\scriptscriptstyle N})
\rangle_{\scriptscriptstyle N}$ over the discrete points
$\mu_{\scriptscriptstyle N}$ in terms of the average $\langle \nct^2 (\mu)
\rangle_\mu$ over the continuous variable $\mu$ around
$\mu_{\scriptscriptstyle N}$ (cf the Appendix),
\begin{equation} \label{112}
\langle \nct^2 (\mu_{\scriptscriptstyle N}) \rangle_{\scriptscriptstyle N} =
\langle \nct^2 (\mu) \rangle_\mu +\frac{1}{6} - \frac{1}{4} \ \rhob^2 \
\langle s_{\scriptscriptstyle N}^2 \rangle_{\scriptscriptstyle N} \ ,
\end{equation}
we get
\begin{equation} \label{104}
\langle \et (N) \rangle_{\scriptscriptstyle N} =
\frac{1}{2 \ \rhob} \ \langle \nct^2 (\mu) \rangle_\mu - \frac{1}{12 \ \rhob}
+ {\cal O} (\mut_{\scriptscriptstyle N}^3) \ .
\end{equation}
The final expression for the average value of the energy in a system
conserving the number of particles is, according to Eqs.\ (\ref{eav}) and
(\ref{104}),
\begin{equation} \label{105}
\langle E (N) \rangle_{\scriptscriptstyle N} = \eb (N) + \frac{1}{2 \ \rhob} \
\langle \nct^2 (\mu) \rangle_\mu - \frac{1}{12 \ \rhob} + {\cal O}
(\mut_{\scriptscriptstyle N}^3) \ .
\end{equation}

It follows from Eq.\ (\ref{105}) that, with respect to the WK or Weyl smooth
terms, the true average of the energy as a function of $N$ is {\em increased}
by a term proportional to the variance of the accumulated level density.
Equation (\ref{105}) contains all the relevant information on the average, and
allows to understand the numerical results presented above. Before making a
quantitative comparison, we first discuss the general aspects involved in that
equation.

\begin{figure}
\includegraphics[width=0.95\columnwidth,clip=true]{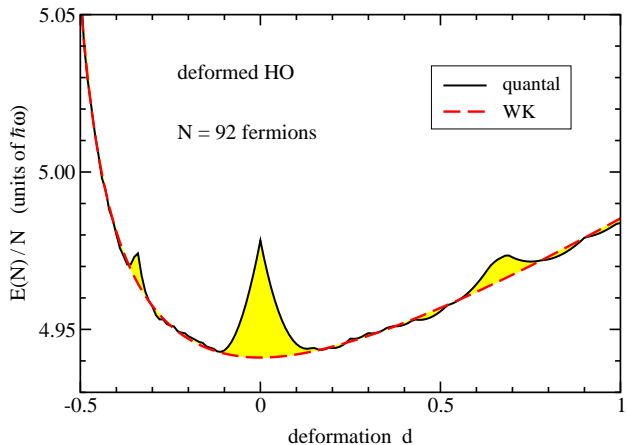}
\caption{\label{figure4} Quantum and WK values of the energy per particle in a
triaxially deformed HO potential. Spin degeneracy is included.}
\end{figure}

Equation (\ref{103}) is demonstrated in the Appendix for a system without
degeneracies (intrinsic and/or due to spin). However, it is easy to see that
it is also valid in the presence of degeneracies. This is because the
thermodynamic quantities we are considering are continuous variables of a
given set of external parameters $\vec{\lambda}$. Assume that for some
$\vec{\lambda} = \vec{\lambda}_0$ degeneracies occur, and that for slightly
different values $\vec{\lambda} \neq \vec{\lambda}_0$ all the degeneracies are
lifted (for instance, some of the components of $\vec{\lambda}$ may be
associated to a shape deformation, with $\vec{\lambda}=\vec{\lambda}_0$ the
spherical case, and another component may be a magnetic field that lifts the
spin degeneracy, with $\vec{\lambda}=\vec{\lambda}_0=0$ no magnetic field).
Then for $\vec{\lambda} \neq \vec{\lambda}_0$ Eq.\ (\ref{103}) is valid. One
can therefore consider the case with degeneracies as the limit $\vec{\lambda}
\rightarrow \vec{\lambda}_0$ and, by continuity, Eq.\ (\ref{103}) remains
valid.

\begin{figure}
\includegraphics[width=0.98\columnwidth,clip=true]{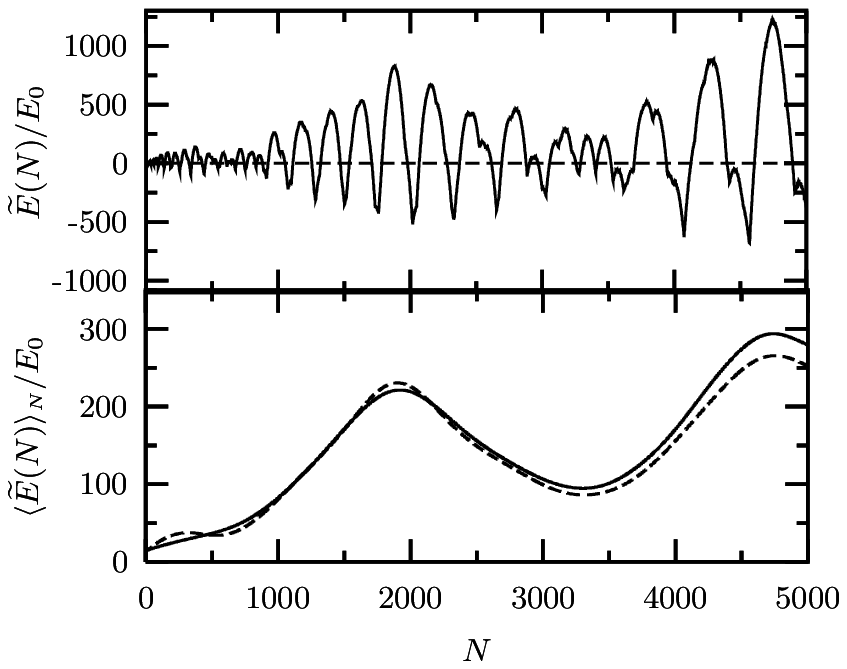}
\caption{\label{figure3} Upper panel: normalized fluctuating part $\et (N)=
E(N) - \eb (N)$ as a function of $N$ for a spherical cavity. $E_0 = \hbar^2/2
m r_0^2$ where $r_0$ is the radius of the sphere and $m$ is the mass of
fermions. Lower panel: average $\langle \et (N) \rangle_{\scriptscriptstyle
N}$ of the fluctuating function in the upper panel (full line) compared to the
theoretical prediction, Eq.\ (\ref{104}) (dashed line).}
\end{figure}

The variance $\langle \nct^2 (\mu) \rangle_\mu$ in Eq.\ (\ref{105}) depends on
the system under consideration. However, its general properties can easily be
determined. In systems where the typical size of the fluctuations is
important, then the shift of the true average $\langle E (N)
\rangle_{\scriptscriptstyle N}$ with respect to $\eb (N)$ will also be
important. On the other hand, in systems with small fluctuations, $\langle
\nct^2 (\mu) \rangle_\mu /2 \rhob$ will be small, and the term $\eb (N)$ will
give not only a good approximation to $\langle E (N)
\rangle_{\scriptscriptstyle N}$, but also to $E (N)$ as well (since
fluctuations are small). In general, the more regular and/or symmetric a
system is, the greater the fluctuations are, and the larger the correction
$\langle \nct^2 (\mu) \rangle_\mu /2 \rhob$ will be. As the regularities or
symmetries are broken, the typical size of the fluctuations diminishes, and $E
(N)$ will be well approximated by $\eb (N)$. This point is illustrated in
Fig.~\ref{figure2}, where the upper panel shows $E (N)/N$ for the isotropic
HO, where large fluctuations are observed and clear deviations of the average
with respect to $\eb (N)$ are found. In contrast, the lower panel shows a
strongly deformed HO, with frequencies $\omega_x/\omega=0.460$,
$\omega_y/\omega=1.111$, and $\omega_z/\omega=1.954$ ($\omega_x \omega_y
\omega_z= \omega^3$), where small fluctuations are observed, and as a
consequence good agreement between $E (N)/N$ and $\eb (N)/N$ is found. Another
manifestation of the same phenomenon is provided in Fig.~\ref{figure4}, where
$E (N)/N$ is shown for $N= 92$ fermions (with spin degeneracy) in a triaxially
deformed HO potential as a function of the deformation parameter $d$, where
$\omega_x/\omega = \delta^{-1/2}/\sigma^{1/3}$, $\omega_y/\omega =
\delta^{1/2}/\sigma^{1/3}$, and $\omega_z/\omega = \sigma^{2/3}$, with
$\sigma=1 + d \sqrt{3}$ and $\delta=1 + \vert d \vert \sqrt{2}$. We see that
for most deformations (mid-shell configurations) the quantum and the smooth WK
values practically agree, up to small fluctuations. Large deviations are
observed, instead, when sphericity is approached, and for other special
deformations, e.g., for $d \sim 0.65$ where the frequency ratio $\omega_x
\!:\! \omega_y \!:\! \omega_z$ is close to $1 \!:\! 2 \!:\! 3$ (when the three
frequencies $\omega_x$, $\omega_y$, and $\omega_z$ are integer ratios, the
energy levels of the HO are degenerate and the classical trajectories of the
Hamiltonian become closed periodic orbits \cite{BB}).

We have made a quantitative check of Eqs.\ (\ref{104}) and (\ref{105}) for the
case of a Fermi gas in a spherical cavity. The upper panel of
Fig.~\ref{figure3} represents the fluctuating part $\et (N)$ as a function of
$N$, defined in Eq.\ (\ref{111}). A clear structure organized in shells (rapid
oscillations) and supershells (long-range modulation of the rapid
oscillations) is observed. The lower panel shows a comparison between the
average $\langle \et (N) \rangle_{\scriptscriptstyle N}$ calculated
numerically from the upper panel of that figure and the result obtained from
Eq.\ (\ref{104}) as a function of $N$. The average shows a nontrivial
dependence with the particle number (which reflects, to a large extent, the
supershell structure), that is very well reproduced by theory.

\begin{figure}
\includegraphics[width=0.98\columnwidth,clip=true]{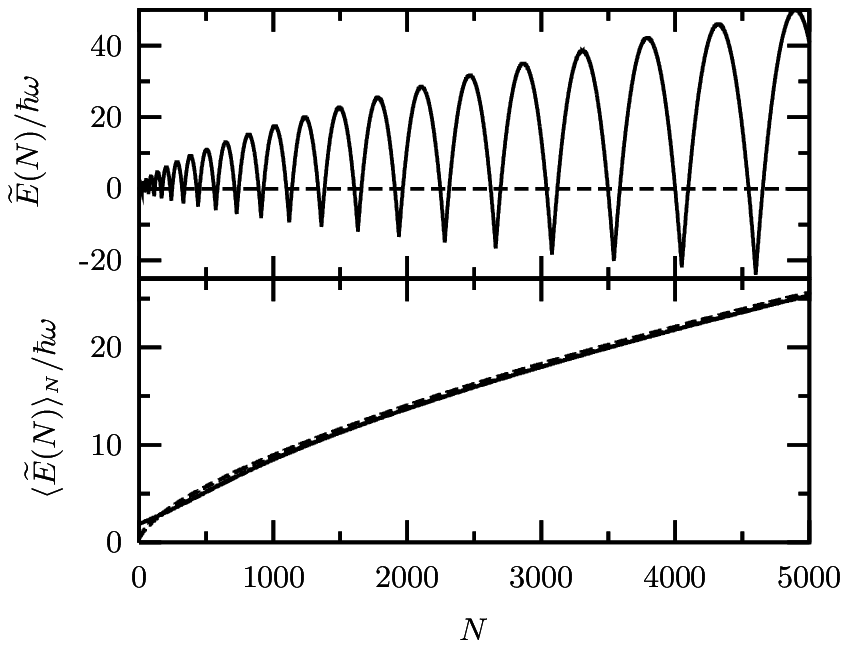}
\caption{\label{figure5} Upper panel: fluctuating part $\et (N)= E (N) -
\eb(N)$ as a function of $N$ for a 3D isotropic HO. Lower panel: average
$\langle \et (N) \rangle_{\scriptscriptstyle N}$ of the fluctuating function
in the upper panel (full line) compared to the theoretical prediction, Eq.\
(\ref{eq30}) (dashed line).}
\end{figure}

In the case of the spherical HO, it is possible to get easily an analytical
expression for $\langle E (N) \rangle_{\scriptscriptstyle N}$. The function
$\nct(\mu)$ is given by the second term in the right hand side of Eq.\
(\ref{eq23}). Squaring it, the main contribution to $\nct^2(\mu)$ comes from
terms where both indices of the double sum are equal. Hence to leading order
in $\mu$, we get:
\begin{equation}
\nct^2(\mu)=\bigg(\frac{\mu}{\hbar \omega}\bigg)^4 \sum_{M=1}^{\infty}
\bigg(\frac{\sin(2 \pi M \mu/\hbar\omega )}{\pi M} \bigg)^2 \ .
\end{equation}
 $\langle \nct^2 (\mu) \rangle_\mu$ is calculated by averaging over the
rapidly fluctuating factors, given by the sine terms. This yields
\begin{equation}
\langle \nct^2 (\mu) \rangle_\mu = \frac{(\mu/\hbar \omega)^4}{2 \pi^2} 
\sum_{M=1}^{\infty} \frac{1}{M^2}
=\frac{1}{12} \bigg( \frac{\mu}{\hbar \omega} \bigg)^4 \ .
\end{equation}
Since $\langle \nct^2 (\mu) \rangle_\mu $ is a smooth function, we replace
$\mu$ by $\mub$. Thus we can use the WK expression Eq.\ (\ref{eq20}) to
compute the dependence of the variance with the number of particles. Using
moreover Eq.\ (\ref{eq19}) we finally get, to leading order in $N$,
\begin{equation}\label{eq30}
\langle \et (N) \rangle_{\scriptscriptstyle N} = \frac{\langle \nct^2 (\mu)
\rangle_\mu}{2 \rhob} = \frac{1}{24} (3 N)^{2/3} \hbar \omega \ .
\end{equation} 
A comparison with the numerical average of $\et(N)$, obtained from an
isotropic 3D HO, is presented in Fig.~\ref{figure5}. The result shows an
excellent agreement; compared to the spherical cavity, a much simpler $N$
dependence is observed, due to the absence of supershells.

Based on general properties of the single-particle spectrum, it is possible to
estimate the typical size of the variance $\langle \nct^2 (\mu) \rangle_\mu$
and of its $N$ dependence for a large class of systems, and therefore to
estimate $\langle E (N) \rangle_{\scriptscriptstyle N}$. The relevant
classification relies on the type of classical dynamics associated to the
confining potential. The two extreme cases that can be treated explicitly are
fully regular and fully chaotic dynamics (the case of mixed dynamics is more
subtle). Based on this classification, explicit results for the typical size
of $\langle \nct^2 (\mu) \rangle_\mu$ were obtained in Ref.~\cite{lm}.

\section{Deformed and self-bound systems}

Up to now we have considered fermion systems confined by an external
potential. This may be applicable to e.g.\ quantum dot systems or magnetically
trapped atomic gases where the self-consistent mean field part plays a minor
role with respect to the external confining potential. However, many relevant
systems are self bound and then the mean field potential is essentially given
by the solution of the self-consistent Hartree-Fock equations which are
obtained by minimising the energy of the system. The mean field in these
situations may turn out to be spherical, but in many cases rotational
invariance will be broken and the mean field becomes deformed. We will see
that in these cases the results can show interesting differences with regard
to the scenario found in the upper panel of Fig.~\ref{figure2} or in
Fig.~\ref{figure5} for the harmonic oscillator, and in Fig.~\ref{figure3} for
the hard wall cavity, where the potential was kept spherical. We want to
investigate such cases now.

First, to illustrate the situation, we again consider the HO potential. In
contrast to the previous section, for each particle number we now minimise the
ground-state energy of the quantum solution with respect to deformation, i.e.,
with respect to free variation of $\omega_x$, $\omega_y$, and $\omega_z$,
under the constraint of volume conservation ($\omega_x \omega_y \omega_z =
\omega^3$). This must be done in carefully checking simultaneously the optimal
choice of the occupancies $n_x, n_y, n_z$. The semiclassical energies $\eb(N)$
always have their absolute minimum at sphericity as given by Eq.\
(\ref{eq21}). As a particular example of self-bound system, we consider the
case of the atomic nuclei. We mimic the saturation properties of nuclear
forces by including the standard particle-number dependence of the HO
frequency $\hbar \omega = 41 A^{-1/3}$ MeV \cite{RS} with $A= 2 N$ (i.e., $A$
here represents the mass number of a hypothetical uncharged nucleus with $N$
protons and $N$ neutrons).

\begin{figure}
\includegraphics[width=0.95\columnwidth,clip=true]{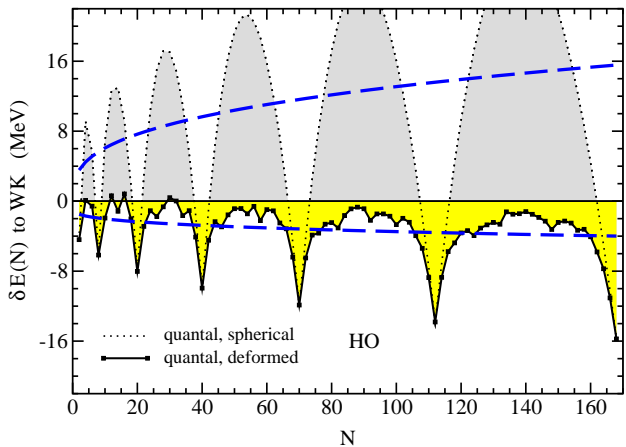}
\caption{\label{figure6} Difference between the minimised ground state energy
with respect to deformations and the spherical WK energy $\eb (N)$ for
fermions in a HO potential (squares joined by full line). For comparison, the
dotted curve shows the energy difference for an isotropic HO (same curve as in
the upper part of Fig.~\ref{figure5}). The average of both curves is shown by
a long-dashed line. The scaling $\hbar \omega = 41 (2 N)^{-1/3}$ MeV has been
used in the calculations.}
\end{figure}

In Fig.~\ref{figure6} we show the difference $\delta E$ between the fully
minimised quantum energies and the corresponding isotropic semiclassical
expression $\eb (N)$ obtained from Eqs.\ (\ref{eq20}) and (\ref{eq21}). For
comparison, we include in the same figure the fluctuating part $\et (N)$ for
the spherical HO (same curve as in the upper part of Fig.~\ref{figure5}). We
observe that with respect to the purely spherical case, the situation changes
very much. Now, contrary to the spherical case, practically all values of
$\delta E$ from the minimised quantum solutions are negative, meaning that the
minimised quantum energies stay {\em below} the semiclassical curve $\eb (N)$.
The minimised quantum energies coincide with the spherical ones only in a
small neighborhood around the shell closures, whereas away from the latter the
system is axially deformed or even, around the middle of the shells, a slight
triaxiality can appear (in the case of axial symmetry, typical deformations
show an axis ratio of 2:3).

It seems natural that the deformed quantum energies are more bound than the
approximate energies obtained from the semiclassical theory, in spite of the
fact that to our knowledge no upper bound theorem like the Rayleigh-Ritz
principle exists for the semiclassical approach. We wish to point out
that in Fig.~\ref{figure6} for most values of $N$ the system is actually
rather well deformed and that, with the exception of a couple of particle
numbers around closed shells, the energy differences are in most cases very
close to the zero line, i.e., to the semiclassical WK values. This is
consistent with the results obtained in the previous section, where it was
shown that for deformed systems where degeneracies are lifted the energy
$E(N)$ is expected to be well approximated by the WK theory.

The magnitude of the difference $\delta E$ of the minimised quantum solutions
to the semiclassical values slightly increases with increasing particle
number, as the average curve shows in Fig.~\ref{figure6}. However, the
magnitude of the same quantity $\delta E$ divided by the particle number
decreases as a function of increasing $N$, and the minimised deformed quantum
energies per particle are extremely close to the semiclassical ones. Notice
the opposite trend in Fig.~\ref{figure6} of the two average curves, with and
without energy minimisation. In contrast to the latter case, for which an
explicit formula for the average behavior was developed and successfully
checked in the previous section, we do not yet have an equivalent result for a
self-bound system.

\begin{figure}
\includegraphics[width=0.95\columnwidth,clip=true]{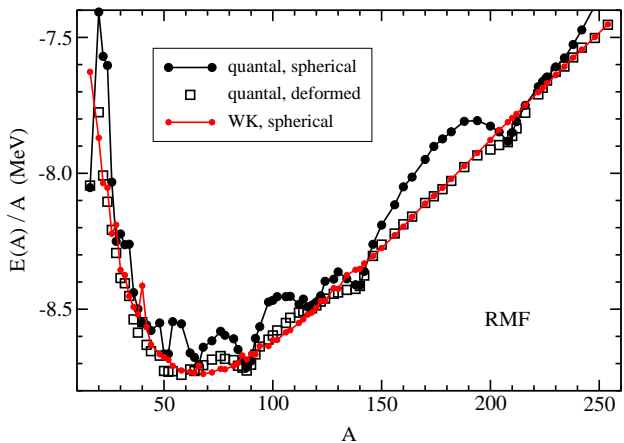}
\caption{\label{figure7} Energy per particle of atomic nuclei with an even
number of protons and of neutrons along the periodic table, as obtained from
self-consistent relativistic mean field calculations. The deformed
calculations are from Ref.~\cite{lalaz99}. For each value of the mass number
$A$ we plot the most bound isotope according to the tabulation of
Ref.~\cite{lalaz99} (excepting for $A=40$ \cite{exception}).}
\end{figure}

We are interested in checking whether this simplified harmonic oscillator
scenario remains valid in realistic Hartree-Fock-type mean field calculations.
In Ref.~\cite{cent06} self-consistent calculations of the ground-state binding
energy of atomic nuclei were carried out using the variational Wigner-Kirkwood
method \cite{cent98}. The nuclear interaction was described by the
relativistic mean field (RMF) meson exchange model \cite{serot86}. Quantum
calculations for the RMF model are available in the literature. In particular,
a mass table of deformed (axially symmetric) quantum calculations for nuclei
with an accurately calibrated RMF nuclear interaction is published in Ref.\
\cite{lalaz99}. From this table we took for each value of the mass number $A$
the {\em most bound} (in general deformed) isotope and traced $E/A$ as a
function of $A$ \cite{exception}. The quantum values together with the RMF
semiclassical results, computed following Ref.~\cite{cent06}, are shown in
Fig.~\ref{figure7} for nuclei with an even number of protons and neutrons.
Most of the WK energies lie on top of the deformed quantum energies on the
scale of the figure. We plot in addition the RMF quantum values constrained to
sphericity. The typical arch structure found in Fig.~\ref{figure2} for the
spherical HO potential is then recovered. These arches in nuclei take place
between the so-called magic numbers, i.e., the proton or neutron numbers where
effects analogous to the shell closures of the HO or of the electron shells in
atoms occur. The fact that for nuclei above iron $E/A$ raises whereas in
Fig.~\ref{figure2} it keeps decreasing is a trivial effect due to the Coulomb
repulsion among protons in the atomic nucleus.

\begin{figure}
\includegraphics[width=0.95\columnwidth,clip=true]{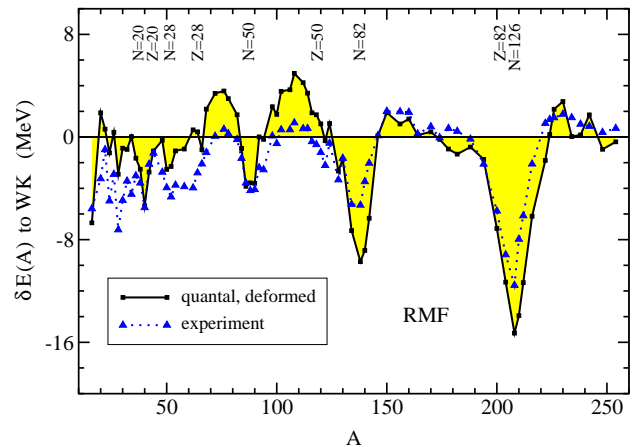} 
\caption{\label{figure8} Energy difference between the deformed quantum
solutions and the WK values of Fig.~7 for the self-consistent relativistic
nuclear mean field. The result obtained by taking the difference of the
experimental data \cite{audi} to the calculated WK energies is also shown for
the purpose of illustration. The location of the magic neutron ($N$) and
proton ($Z$) numbers is indicated.}
\end{figure}

In Fig.~\ref{figure8} we display for the self-consistent RMF the difference
$\delta E$ between the quantum energies (that are, as mentioned, minimised
with respect to deformation) and the corresponding semiclassical values (that
attain their absolute minima at sphericity). For reference, also the values of
the energy difference $\delta E$ obtained from experimental data \cite{audi}
instead of the quantum results are displayed. The similarity of
Fig.~\ref{figure8} with Fig.~\ref{figure6} for the HO is striking. Systems
with the largest deformations are again located mostly around mid shells. When
the system approaches the spherical shape, $\delta E$ becomes increasingly
negative and displays the downward peaks seen in Fig.~\ref{figure8} on
reaching neutron or proton magic numbers.

It is clear that in the self-consistent case as in the schematic case of the
HO with optimized shapes considered above, the average of $\delta E$ as a
function of particle number is, at least for the heavier systems, negative. In
self-bound systems we again find that in between shells the quantum energies
are closer to the semiclassical WK values. In between shells the system
deforms in search of minimum energy and avoids the large positive shell
corrections to the energy that occur if a spherical shape is kept. As the
deformation increases, symmetries are broken and the amplitude of the shell
corrections diminishes. This is in agreement with the basic ideas underlying
Eq.\ (\ref{105}), that imply an energy $E (N)$ which is well approximated by
the WK theory.

\section{Concluding remarks}

In this work we have revisited the old problem of the semiclassical approach
to finite fermion systems, based either on the Wigner-Kirkwood expansion with
$\hbar$ corrections, or the Weyl expansion. We have addressed the nature of
one of the most elusive features of the theory, namely the systematic
overbinding compared to the quantum average for fermions in fixed potentials.

In the first part, we have shown that this discrepancy is due to the fact that
these methods do not incorporate appropriately the conservation of the
particle number. There is, generically, a contribution to the average
ground-state energy that comes from the fluctuating part, or shell
contribution. We derived an explicit formula that takes into account that
contribution, and have tested it for different fixed confining potentials. In
all cases, a positive correction with respect to the semiclassical result is
predicted [cf Eq.\ (\ref{105})], whose magnitude depends on the size of the
shell effects. When the confining potential has symmetries, the shell
corrections are large, and important deviations between the exact quantum and
the WK energies are observed, in agreement with our predictions. In contrast,
when symmetries are broken shell effects are smaller, and the exact energies
(and not only their average part) are better described by the WK theory.

The description of the behavior of self-bound systems is more difficult and
subtle, because at each particle number $N$ the energy is minimised, and hence
the shape of the potential is a function of $N$. In this case, a shell
correction which is nearly always negative with respect to the spherical WK
result is observed for the HO potential. In between shells, when the system
deforms and symmetries are broken, the value of the shell correction is
smaller, and the energy is well approximated by the WK theory, in agreement
with the general considerations that follow from Eq.\ (\ref{105}).
Interestingly, all these features have been qualitatively confirmed by a more
realistic model based on a mean-field self-consistent calculation of the
ground-state energy of atomic nuclei. However, the problem of deriving an
explicit formula for the average behavior of the ground-state energy of
self-bound systems that correctly takes into account the $N$-dependence with
deformation degrees of freedom is still open.

We are indebted to R. K. Bhaduri and O. Bohigas for valuable informations.
Work partially supported by IN2P3-CICYT\@. X.V. and M.C. acknowledge Grants
No.\ FIS2005-03142 (MEC, Spain, and FEDER) and No.\ 2005SGR-00343 (Generalitat
de Catalunya). P. L. and J. R. acknowledge support from grants ACI Nanoscience
201, ANR NT05-2-42103, ANR-05-Nano-008-02 and the IFRAF Institute.

\renewcommand{\theequation}{A\arabic{equation}}
\setcounter{equation}{0}   
\section*{APPENDIX}

We follow here Appendix B of Ref.~\cite{Mon} in order to prove
Eqs.\ (\ref{103}) and (\ref{112}).

Let us consider a single-particle spectrum $\varepsilon_j$, with
$j=1,2,\ldots$ and $\varepsilon_j \leq \varepsilon_{j+1}$. The accumulated
level density $\nc (\mu)$ is discontinuous at each energy level. At the
discontinuity, we assign to $\nc (\mu)$ the ``intermediate'' value $\nc
(\varepsilon_{\scriptscriptstyle N} ) = N -1/2$. For $N \gg 1$ and
$\varepsilon_{\scriptscriptstyle N} < \mu < \varepsilon_{\scriptscriptstyle
N+1}$, writing $\nc (\mu) = \ncb (\mu) + \nct (\mu)$, making a Taylor
expansion of $\ncb (\mu)$ around $\varepsilon_{\scriptscriptstyle N}$, and
using that $\nc (\varepsilon_{\scriptscriptstyle N} ) = N -1/2$, we may write
\begin{equation} \label{A1}
\nct (\mu) = \nct (\varepsilon_{\scriptscriptstyle N} ) + \frac{1}{2} - (\mu -
\varepsilon_{\scriptscriptstyle N} ) \rhob \ .
\end{equation}
Evaluating this relation just before $\mu=\varepsilon_{\scriptscriptstyle
N+1}$, and taking into account the value of the function at
$\varepsilon_{\scriptscriptstyle N+1}$, we have
\begin{equation} \label{A2}
\nct (\varepsilon_{\scriptscriptstyle N+1}) = \nct
(\varepsilon_{\scriptscriptstyle N} ) + 1 - s_{\scriptscriptstyle N} \rhob \ ,
\end{equation}
where 
$$
s_{\scriptscriptstyle N} = \varepsilon_{\scriptscriptstyle N+1} -
\varepsilon_{\scriptscriptstyle N}
$$ 
is the level spacing (we neglect the dependence of $\rhob$ with energy).
Taking the discrete average over $N$ on both sides of Eq.\ (\ref{A2}), defined
as
$$ 
\langle f(\varepsilon_{\scriptscriptstyle N} ) \rangle_{\scriptscriptstyle N} 
= \frac{1}{\Delta N} \sum_{j=N-\Delta N/2}^{N+\Delta N/2} f (\varepsilon_j) \ ,
$$ 
where $\Delta N$ is the number of levels in the window around the $N$th level,
and using that $\langle \nct (\varepsilon_{\scriptscriptstyle N+1})
\rangle_{\scriptscriptstyle N} = \langle \nct (\varepsilon_{\scriptscriptstyle
N} ) \rangle_{\scriptscriptstyle N}$, we obtain the (trivial) relation
$\langle s_{\scriptscriptstyle N} \rangle_{\scriptscriptstyle N} = 1/\rhob$.

On the other hand, defining the continuous average over a window $\Delta \mu$
of a function that depends on the chemical potential as
$$ 
\langle f(\mu) \rangle_\mu = \frac{1}{\Delta \mu} \sum_{j=N-\Delta
N/2}^{N+\Delta N/2} \int_{\varepsilon_j}^{\varepsilon_{j+1}} f (\mu) \ d \mu \
,
$$ 
where $\Delta N = \rhob \Delta \mu$ is the number of levels in the window,
we have from Eq.\ (\ref{A1})
\begin{eqnarray}
&& \langle \nct (\mu) \rangle_{\mu} = \nonumber \\ &=& \frac{1}{\Delta \mu}
\sum_{j=N-\Delta N/2}^{N+\Delta N/2} \int_{\varepsilon_j}^{\varepsilon_{j+1}}
\left[ \nct (\varepsilon_j) + 1/2 - \rhob (\mu - \varepsilon_j ) \right] \ d
\mu \nonumber \\ &=& \frac{\Delta N}{\Delta \mu} \left( \langle \nct
(\varepsilon_{\scriptscriptstyle N} ) s_{\scriptscriptstyle N}
\rangle_{\scriptscriptstyle N} + \frac{\langle s_{\scriptscriptstyle N}
\rangle_{\scriptscriptstyle N}}{2} - \frac{\rhob}{2} \langle
s_{\scriptscriptstyle N}^2 \rangle_{\scriptscriptstyle N} \right) \nonumber \\
&=& \rhob \left( \langle \nct (\varepsilon_{\scriptscriptstyle N} )
s_{\scriptscriptstyle N} \rangle_{\scriptscriptstyle N} + \frac{1}{2 \rhob} -
\frac{\rhob}{2} \langle s_{\scriptscriptstyle N}^2 \rangle_{\scriptscriptstyle
N} \right) = 0 \ .
\label{A3}
\end{eqnarray}
The last equality follows because $\langle \nct (\mu) \rangle_{\mu}=0$
by definition. From  Eq.\ (\ref{A3}) we deduce
\begin{equation} \label{A3b}
\langle \nct (\varepsilon_{\scriptscriptstyle N} ) s_{\scriptscriptstyle N}
\rangle_{\scriptscriptstyle N} = \frac{\rhob}{2} \langle s_{\scriptscriptstyle
N}^2 \rangle_{\scriptscriptstyle N} - \frac{1}{2 \rhob} \ .
\end{equation}
Squaring and computing the discrete average in both sides of Eq.~(\ref{A2}) it
is possible to deduce that $\langle \nct (\varepsilon_{\scriptscriptstyle N} )
\rangle_{\scriptscriptstyle N} = 0$ after using Eq.~(\ref{A3b}).

Similarly, squaring Eq.~(\ref{A1}) and computing in both sides the continuous
average it is possible to relate the continuous variance of $\nct (\mu)$ with
discrete averages
\begin{equation} \label{A4}
\langle \nct^2 (\mu) \rangle_{\mu} = \rhob \langle \nct^2
(\varepsilon_{\scriptscriptstyle N} ) s_{\scriptscriptstyle N}
\rangle_{\scriptscriptstyle N} - \rhob^2 \langle \nct
(\varepsilon_{\scriptscriptstyle N} ) s_{\scriptscriptstyle N}^2
\rangle_{\scriptscriptstyle N} + \frac{\rhob}{3} \langle s_{\scriptscriptstyle
N}^3 \rangle_{\scriptscriptstyle N} - \frac{1}{4} \ .
\end{equation}
Computing the discrete average of the third power of Eq.~(\ref{A2}), the
previous expression for the continuous variance is considerably simplified and
gives
\begin{equation} \label{A5}
\langle \nct^2 (\varepsilon_{N}) \rangle_{\scriptscriptstyle N} = \langle
\nct^2 (\mu) \rangle_{\mu} - \frac{1}{12} \ .
\end{equation}

We are now interested in the statistics at the discrete points
$\mu_{\scriptscriptstyle N} = (\varepsilon_{\scriptscriptstyle N+1} +
\varepsilon_{N})/2$. From Eq.~(\ref{A1}) we have
\begin{equation} \label{A6}
\nct (\mu_{\scriptscriptstyle N}) = \nct (\varepsilon_{N}) + \frac{1}{2} -
\frac{\rhob}{2} s_{\scriptscriptstyle N} \ ,
\end{equation}
from which it is easy to deduce that $\langle \nct (\mu_{\scriptscriptstyle N}
) \rangle_{\scriptscriptstyle N} = 0$. From the discrete average of the square
of Eq.~(\ref{A6}), using the result (\ref{A3b}) as well as Eq.~(\ref{A5}), it
follows that
\begin{equation} \label{A7}
\langle \nct^2 (\mu_{\scriptscriptstyle N}) \rangle_{\scriptscriptstyle N} =
\langle \nct^2 (\mu) \rangle_{\mu} + \frac{1}{6} - \frac{\rhob^2}{4} \langle
s_{\scriptscriptstyle N}^2 \rangle_{\scriptscriptstyle N} \ .
\end{equation}
We have demonstrated Eq.~(\ref{112}).

Let us now consider the grand potential for $N \gg 1$ and
$\varepsilon_{\scriptscriptstyle N} < \mu < \varepsilon_{\scriptscriptstyle
N+1}$. Since $\ot (\mu) - \ot (\varepsilon_{\scriptscriptstyle N} ) = -
\int_{\varepsilon_{\scriptscriptstyle N}}^{\mu} \nct (\varepsilon) \ d
\varepsilon$, using Eq.~(\ref{A1}) and integrating we get
\begin{equation} \label{A8}
\ot (\mu) = \ot (\varepsilon_{\scriptscriptstyle N} ) - \left( \nct
(\varepsilon_{\scriptscriptstyle N} ) + \frac{1}{2} \right) (\mu -
\varepsilon_{\scriptscriptstyle N} ) + \frac{\rhob}{2} (\mu -
\varepsilon_{\scriptscriptstyle N} )^2 \ .
\end{equation}
Noting that $\ot (\mu)$ is a continuous function, Eq.\ (\ref{A8}) at
$\mu=\varepsilon_{\scriptscriptstyle N+1}$ gives
\begin{equation} \label{A9}
\ot (\varepsilon_{\scriptscriptstyle N+1}) = \ot
(\varepsilon_{\scriptscriptstyle N} ) - \left( \nct
(\varepsilon_{\scriptscriptstyle N} ) + \frac{1}{2} \right)
s_{\scriptscriptstyle N} + \frac{\rhob}{2} s_{\scriptscriptstyle N}^2 \ .
\end{equation}
In a similar way as it was done for the accumulated level density, by
integration of Eq.~(\ref{A8}) (knowing that $\langle \ot (\mu) \rangle_{\mu} =
0$), and discrete average of the product of Eq.~(\ref{A2}) and Eq.~(\ref{A9}),
we deduce
\begin{equation} \label{A10}
\langle \ot (\varepsilon_{\scriptscriptstyle N}) \rangle_{\scriptscriptstyle
N} = \frac{1}{\rhob} \langle \nct^2 (\mu) \rangle_{\mu} \ .
\end{equation}
From Eq.~(\ref{A8}), for $\mu = \mu_{\scriptscriptstyle N}$, we have
\begin{equation} \label{A11}
\ot (\mu_{\scriptscriptstyle N}) = \ot (\varepsilon_{\scriptscriptstyle N} ) -
 \left( \nct (\varepsilon_{\scriptscriptstyle N} ) + \frac{1}{2} \right)
 \frac{s_{\scriptscriptstyle N}}{2} + \frac{\rhob}{8} s_{\scriptscriptstyle
 N}^2 \ .
\end{equation}
The discrete average of this equation, together with Eqs.~~(\ref{A3b}) and
(\ref{A10}), finally leads to
\begin{equation} \label{A12}
\langle \ot (\mu_{\scriptscriptstyle N}) \rangle_{\scriptscriptstyle N} =
\frac{1}{\rhob} \langle \nct^2 (\mu) \rangle_{\mu} - \frac{\rhob}{8} \langle
s_{\scriptscriptstyle N}^2 \rangle_{\scriptscriptstyle N} \ .
\end{equation}
This equation, together with Eq.\ (\ref{A7}), implies Eq.~(\ref{103}).

%



\begin{thebibliography}{99}
\bibitem{footnote} We use the notation $E
= \eb + \et$, $\nc= \ncb + \nct$, etc.\ to indicate the total, average, and
fluctuating parts of each magnitude. However, in the axis labels of the
figures where different approximations are compared, we shall write only $E$,
$\nc$, etc.
\bibitem{exception}
We have made an exception to the rule of selecting the most bound
isotope for each mass number in the case of $A=40$. The
most bound $A=40$ nucleus is $^{40}$Ar ($Z=18,N=22$). However, in
Figs.~\ref{figure7} and \ref{figure8} we display the results for
$^{40}$Ca instead of $^{40}$Ar, because $^{40}$Ca is one of the scarce
examples of a stable nucleus with a double shell closure ($Z=N=20$).
\end{thebibliography}
\end{document}